\documentclass[11pt]{emulateapj}\usepackage{times}

\usepackage[usenames]{color}

\usepackage{graphicx}
\usepackage{epstopdf}
\usepackage{url}

\def\aa {1E\,1547$-$5408}

\def\xte{XTE\,J1810$-$197\,}

\def\sgra{SGR\,1806$-$20}

\def\sgrd{SGR\,1627$-$41\,}

\def\psr{PSR\,1622$-$4950}
\def\hbpsr{PSR\,J1846$-$0258}

\def\ergs {erg\,s$^{-1}$}
\def\ergscm2 {erg\,s$^{-1}$cm$^{-2}$}

\def\cm2 {cm$^{-2}$}

\begin{document}

\shorttitle{\textsc{The fundamental plane for radio magnetars}}
\shortauthors{\textsc{Rea et al.}}

\title{\textsc{The fundamental plane for radio magnetars}}

\author{Nanda Rea\altaffilmark{1},  Jos\'e  A. Pons\altaffilmark{2}, Diego F. Torres\altaffilmark{1,3} \& Roberto Turolla\altaffilmark{4,5}
}

\altaffiltext{1}{Institut de Ci\`encies de l'Espai (CSIC--IEEC),              
              Campus UAB,  Torre C5, 2a planta,
              08193 Barcelona, Spain.}
\altaffiltext{2}{Departament de Fisica Aplicada, Universitat d'Alacant, Ap. Correus 99, 03080 Alacant, Spain.} 
\altaffiltext{3}{Instituci\'o Catalana de Recerca i Estudis Avan\c{c}ats (ICREA), Barcelona, Spain.}
\altaffiltext{4}{Universit\'a di Padova, Dipartimento di Fisica, via F. Marzolo 8, I-35131 Padova, Italy.}
\altaffiltext{5}{Mullard Space Science Laboratory, University College London, Dorking, Surrey RH5 6NT.}

\begin{abstract}
High magnetic fields are a distinguishing feature of neutron stars
and the existence of sources (the soft gamma repeaters and
the anomalous X-ray pulsars) hosting an ultra-magnetized
neutron star (or magnetar) has been recognized in the past few
decades. Magnetars are believed to be powered by magnetic energy
and not by rotation, as with normal radio pulsars. Until recently,
the radio quietness and
magnetic fields typically above the quantum critical value ($B_{\rm Q}\simeq4.4\times10^{13}$\,G), were among the characterizing properties of magnetars. The
recent discovery of radio pulsed emission from a few of them, and of a
low dipolar magnetic field soft gamma repeater, weakened further the idea of a clean separation between
normal pulsars and magnetars. In this Letter we show that radio emission from magnetars might be  powered by rotational energy, similarly to what occurs in normal radio pulsars. The peculiar characteristics of magnetars radio emission should be traced in the complex magnetic geometry of these sources. Furthermore, we propose that magnetar radio activity or inactivity can be predicted from the knowledge of the star's rotational period,
its time derivative and the quiescent X-ray luminosity.
\end{abstract}


\section{Introduction}

The radio emission from pulsars has been studied in detail in
the past decades, and it is believed to be powered by the
star's rotational energy ($L_{\rm rot}= 4\pi^2 I \dot{P}/P^3 \sim 3.9\times10^{49} \dot{P}/P^3$ \ergs, where
$I\sim10^{45}$~g\,cm$^2$ is the star moment of inertia, $P$ and $\dot{P}$ the rotational period (in seconds)
and period derivative). A key ingredient to activate the radio
emission is the acceleration of charged particles, which are
extracted from the star's surface by an electrical voltage gap. The
voltage gap forms due to the presence of a (mainly) dipolar
magnetic field co-rotating with the pulsar, and extends up to an
altitude of $\approx 10^{4}$\,cm with a potential difference $>
10^{10}$\,statvolts . Primary charges are accelerated by the electric
field along the magnetic field lines to relativistic speeds and
emit curvature radiation. Curvature photons are then converted
into electron-positron pairs in the strong magnetic field and this
eventually leads to a pair cascade which is ultimately responsible
for the coherent radio emission we observe from radio pulsars
\citep{gj69,rs75}

\noindent
Under the usual
assumption of dipole magneto-rotational braking, the polar
magnetic field of neutron stars is given by

\begin{equation}
B_{\rm p}  = (3~I~c^3 \dot{P}~P/2\pi^2 R^6)^{1/2} \sim 6.4\times10^{19}\sqrt{P \dot{P}}~{\rm G},
\end{equation}

\noindent where $R\sim10^6$~cm is the neutron
star radius. On the other hand, the electric potential of
pulsars (for slow rotators) can be approximated as
\citep{rs75}

\begin{equation}
\Delta V  = 2~\pi^2~B_{\rm p}~R^3/c^2~P^2 \sim 4.2\times 10^{20}\sqrt{\dot P/P^3} ~{\rm statV}\,.
\end{equation}

\noindent
In the last two decades two new classes of pulsars
were discovered, with properties much at
variance with those of radio pulsars, namely the soft gamma
repeaters (SGR) and the anomalous X-ray pulsars (AXP). SGRs were
discovered as flaring sources by large field-of-view X-ray and
gamma-ray instruments. AXPs, instead,  were observed as steady,
radio-quiet X-ray sources with slow rotational periods (a few
seconds). Despite their apparently different properties, the
detection of powerful bursts from AXPs \citep{gavriil02}, as well as the discovery
of persistent X-ray emission and slow rotational periods from
SGRs \citep{kouv98}, has by now highlighted the connections among the two groups,
and was instrumental in establishing their common magnetar nature
\citep{td95,td01,mereghetti08}.

Until recently, the defining observational properties of magnetars
were: 1- lack of radio emission, possibly due to the quenching of
the pair cascade by photon splitting in their super-strong field
\citep{bh98}; 2- persistent X-ray luminosities exceeding
rotational energy losses ($L_{\rm x}> L_{\rm rot}$), clearly
indicating that these sources are powered by magnetic energy rather than rotation
\citep{td95}; 3- non-thermal emission in the X-ray spectrum,
interpreted as produced by resonant cyclotron upscattering  in the
current-loaded twisted magnetosphere \citep{tlk02}; and 4-
surface dipolar magnetic fields exceeding the quantum critical
value for electrons ($B_{\rm Q}=m_e^2c^3/e\hbar\sim
4.4\times10^{13}$\,G ).


\begin{figure}
\includegraphics[width=0.5\textwidth]{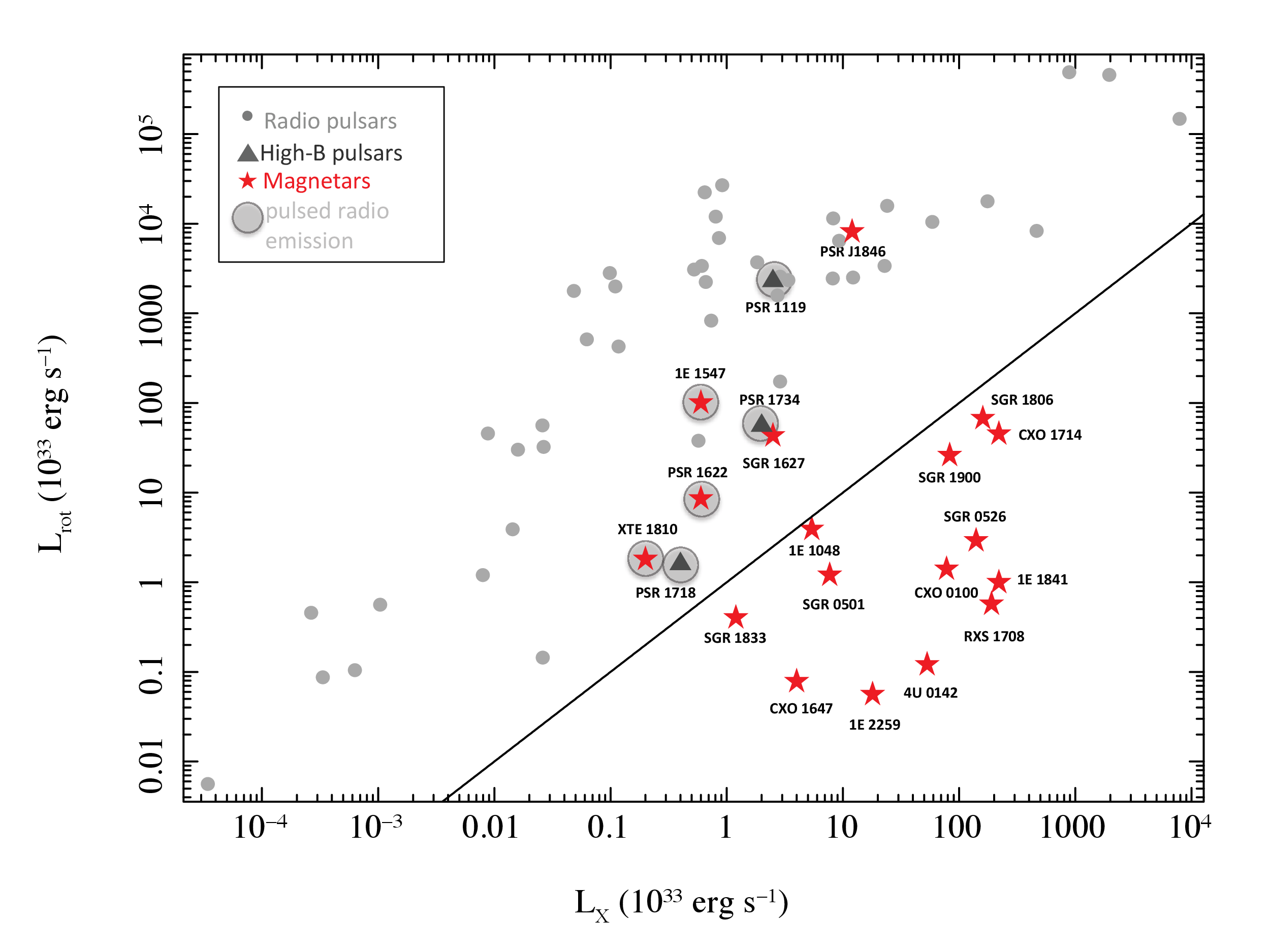}
\caption{X-ray luminosity versus the spin-down luminosity for all pulsars having a detected X-ray emission (grey filled circles), high-B pulsars (filled triangle), and the magnetars (red stars). Grey shaded circles mark the magnetars and high-B pulsars with detected pulsed radio emission, and the solid line shows $L_{\rm x} = L_{\rm rot}$. X-ray luminosities are calculated in the 0.5--10\,keV energy range, and for variable sources refer to the quiescent emission state. }
\label{fig1}
\end{figure}


SGRs and AXPs are now known to exhibit transient X-ray and gamma-ray
outbursts on timescales of months to years, during which an energy
up to $\sim10^{44}$~erg is released. Since the discovery in 2003
of such transient outbursts  \citep{ibrahim04}, 5  new magnetars (over a
total of about 20 confirmed class members) have been discovered by means of their X-ray transient activity and thanks to the current X-ray all-sky  monitors  (the Swift and Fermi satellites; see \cite{re11} for a recent review).

With the discovery of transient magnetars many of the previously
established properties of SGRs/AXPs had to be revised: 1- they can
be radio-loud, though also transient in their radio emission \citep{camilo06}; 2- during the quiescent
state their spin-down luminosity can exceed the X-ray luminosity; 3-
they can have purely thermal spectra; and 4- their surface dipolar magnetic field
can be as low as a few times $10^{12}$\,G, in line with
rotation-powered pulsars \citep{rea10}. In light of this, the
idea that the physics involved in these sources is completely set
apart from that of normal radio pulsars became arguable, as already hinted by the discovery of radio pulsars with magnetic fields reaching into the magnetar range \citep{camilo00,mclaughlin03}. However,
the exact extent of the connection between radio pulsars and
magnetars has been so far a matter of debate. In this respect, the
understanding of magnetar radio emission is crucial in obtaining
a complete picture of the neutron star population as a whole.

\section{Radio emission from magnetars}
\label{magnetars}

The detection of pulsed radio emission from the magnetar \xte\
(Camilo et al. 2006; $B_{\rm p}\sim2.1\times10^{14}$\,G) opened a new
perspective in the study of such strongly magnetized sources, and
the physics of their magnetosphere. For many months,
\xte\ was found to be the strongest radio pulsar in the Galaxy at
frequencies above 20~GHz.  Its radio emission was highly variable
in intensity and pulse-profile morphology on several timescales, and it likely started
around a year after the X-ray outburst onset and then declined in a few years \citep{camilo06,camilo07a,lazaridis08,serylak09}.

Pulsed radio emission was  later discovered to follow the X-ray outbursts
of the magnetar \aa\, (Camilo et al. 2007b; $B_{\rm p}\sim2.2\times10^{14}$\,G). This
source showed three X-ray outbursts in the past 5 years. Between
the last two events, radio emission was observed to decline, and rise again
after the onset of the subsequent X-ray outburst, with a delay of
at least a few days \citep{camilo09,burgay09}.

Very recently yet another radio-pulsed magnetar has been
discovered. \psr\ ($B_{\rm p}\sim2.8\times10^{14}$\,G) was the
first magnetar discovered in the radio band,  with the
identification of its X-ray counterpart following later
\citep{levin10}. In this case the peak of the X-ray
outburst was probably missed \citep{anderson11}, and its
dim X-ray emission is currently fading off, as expected from a
magnetar returning to its quiescent state.

Beside the magnetars reported above, no
other source of the class has shown evidence of radio activity
\citep{burgay06,crawford07,lazarus11}. 

The main properties of the radio-pulsed emission of these sources
are: 1- a delay in the appearance of the radio emission after the
X-ray outburst onset; 2- variable pulse profiles and radio flux on
a timescale from minutes to days; 3- decay of the average radio flux
as the X-ray outburst decays; 4- flat radio spectrum over a wide
range of frequencies (spectral index $\alpha\sim 0$). 
These characteristics appear at variance with those of the radio pulsars having large magnetic fields ($B_{p}\sim 5-9\times10^{13}$\,G; Camilo et al. 2000; McLaughlin et al. 2003; Ng \& Kaspi 2011), which have radio properties in line with those of typical rotation-powered pulsars (i.e. more stable pulse profile
morphology, steep radio spectra, and long-term flux stability).

No complete theory for the ephemeral radio-pulsed emission observed
in outbursting magnetars has been put forward so far, although a few theoretical works started to address this issue \citep{belo09,thompson08b}. Large
observational efforts are on-going to understand when and which
magnetar will emit radio-pulsed emission, or be radio-quiet. So
far it was argued that whatever the mechanism is, it should differ from that of rotation-powered radio pulsars, and that any
magnetar undergoing an outburst could in principle emit radio
waves.  In the following we show that this might not be the case, and only magnetars with certain characteristics would show radio pulsed emission.


\begin{figure*}
\hspace{-0.5cm}
\includegraphics[width=0.54\textwidth]{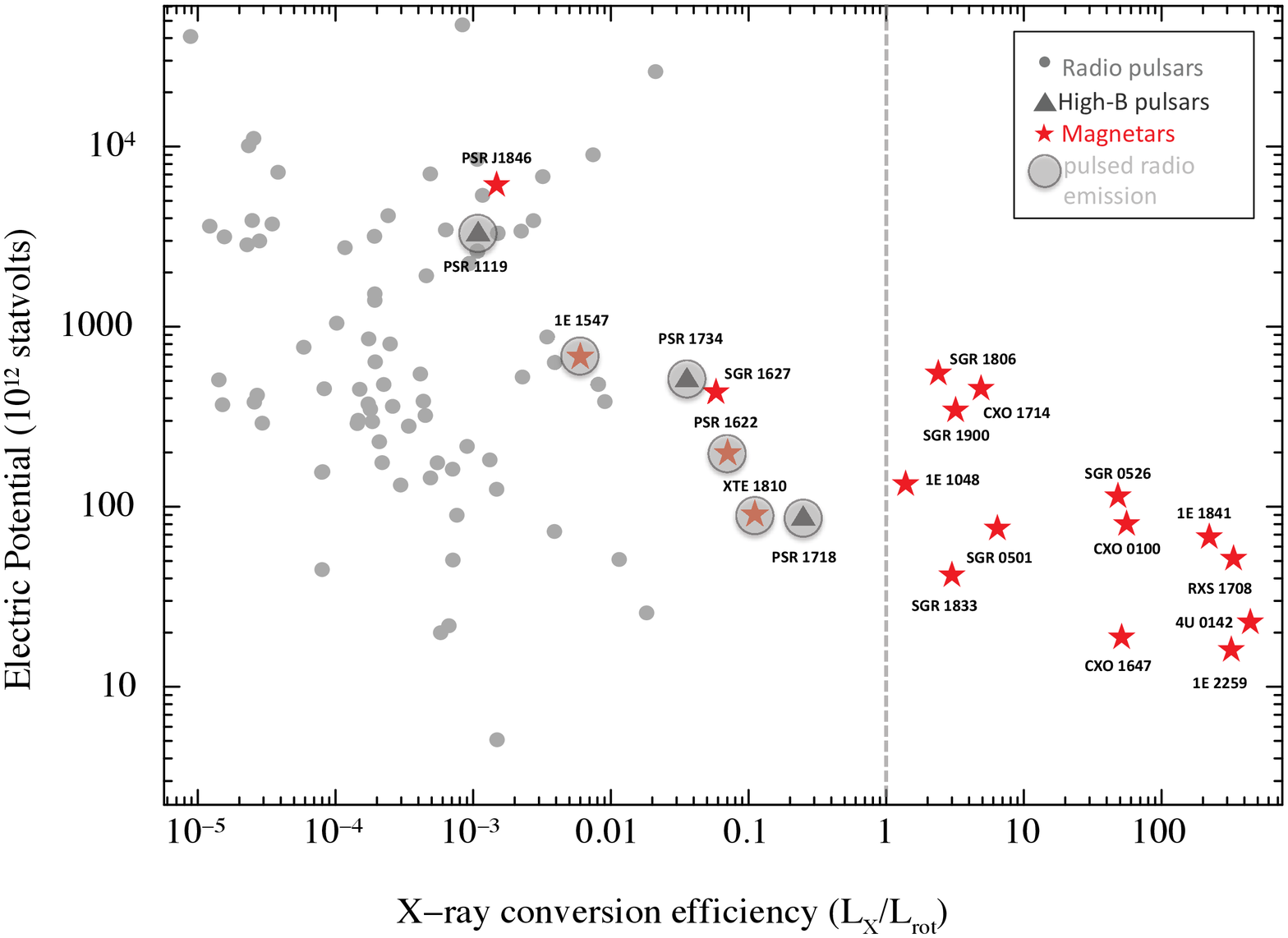}
\hspace{-0.6cm}
\includegraphics[width=0.5\textwidth]{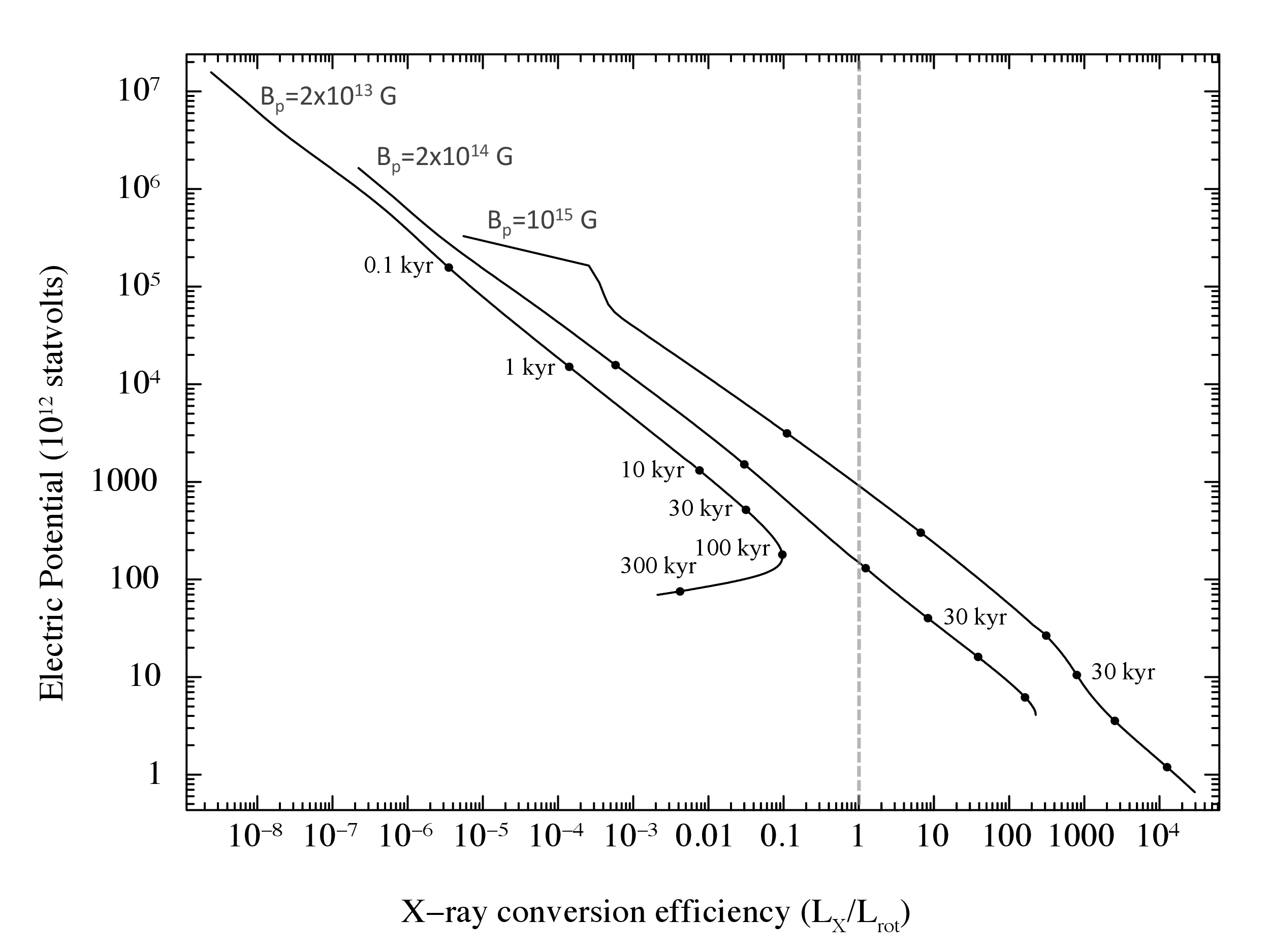}
\caption{{\em Left panel}: {\it Fundamental plane for radio
magnetars}: Electric potential gap as a function of the $L_{\rm x}/L_{\rm rot}$ ratio
for the same sources as in Figure 1. {\em Right panel}: Evolution of the potential gap versus the X-ray conversion efficiency for neutron stars with three initial magnetic field values. The time-steps superimposed to the evolution lines are the same for all lines. See text for details. }
\label{fig1}
\end{figure*}


\section{Results}
\label{results}

In Figure 1 we plot  the quiescent X-ray luminosity (in the 0.5--10 keV energy range) versus the spin-down luminosity, for all known magnetars \citep{re11}, X-ray bright high-B radio pulsars \citep{nk11} and X-ray-emitting rotation-powered radio pulsars \citep{becker09}. The first interesting feature which emerges from this
comparison is that all radio emitting magnetars have a spin-down luminosity larger than their X-ray luminosity, in line with the rotational powered radio
pulsars. In particular,  radio magnetars and high-B radio pulsars tend to fill the gap between normal pulsars and canonical magnetars. 

In Figure 2 (left) we plot the electric potential gap (as from Eq. 2) versus the X-ray efficiency $L_{\rm x}/L_{\rm rot}$,  defined as the ratio of the
quiescent X-ray luminosity (in the 0.5--10 keV energy range) to the spin-down luminosity, for the same sources as for Figure 1. The X-ray conversion efficiency  has been interpreted as the capacity of the pulsar in converting rotational energy into X-ray emission \citep{possenti02,vby11}. For radio-quiet magnetars, it strongly suggests that spin-down luminosity cannot be the "main" responsible of the X-ray emission, which is in fact likely dominated by magnetic energy. On the other hand, the value of the potential gap relates to the ability of the pulsar in extracting and accelerating particles in the polar cap to power the cascade process eventually responsible for the radio emission. 

From Figure 2 (left) it is clear that the potential gap of magnetars is in line with that of rotational powered pulsars, although a decay trend is visible (see below for further details).

We note that two objects, \hbpsr\, and \sgrd, appear at first sight as possible outliers in the otherwise clear trend shown in Figure\,1 and 2. In principle, one could appeal to beaming effects to explain why these 2 magnetars are not detected in radio (among those with $L_{\rm x}/L_{\rm rot}<1$).
However, we note that so far all radio emitting magnetars (for which X-ray pulsations are detected) showed a good X and radio phase-alignment and broad profiles
in the two bands (0.2--0.5 in phase; Serylak et al. 2009). Hence, assuming that all magnetars share similar X/radio properties, there would be good chances for the radio beam to be observable since X-ray pulsations are clearly detected. We then rather think that there are other (more likely) alternatives to explain the current non-detection in radio.

\hbpsr\, is the youngest pulsar (900 yrs) that showed magnetar-like activity  \citep{gavriil08,samar08}.  Deep radio searches were performed during its 2006 magnetar-like outburst and about 18
months later. If the radio emission of \hbpsr\ parallels that of
\xte, the source should have become radio bright sometime after
the X-ray outburst and then continue to fade, reaching in 18 months a flux lower than the $\sim5\ \mu$Jy upper limit (at
1.9 GHz) derived by current observations. If its radio emission was similar to
those of other radio magnetars, current deep radio observations
would have probably missed it \citep{archibald08}. The source distance is still uncertain, ranging from 5 to 21\,kpc (Becker et al. 1984; Leahy \& Tian 2008), which reflects in a large uncertainty in the source dispersion measure (DM); the entire Galactic DM in this direction is $\sim1470$\,pc~cm$^{-3}$ (using the NE2001 model; Cordes \& Lazio 2002). Furthermore, this pulsar lies in a very dense supernova remnant (Kes 75). In such an environment the detection of any radio-pulsed emission is very difficult. 

Searches
for radio emission from \sgrd\ have been recently performed
following its 2008 X-ray outburst \citep{cs08,esposito09}. No radio emission has been
detected, but the large distance ($\sim$11 kpc), large column
density along the line of sight ($N_{\rm H}\sim10^{23}$ \cm2 ; which corresponds to a DM$\sim 1150$\,pc~cm$^{-3}$ using the NE2001 model),
and the time of the observations (close to the outburst peak) make
the non-detection quite un-constraining. We conclude that the two
apparent outliers to the general trend depicted in Figure 1 and 2, have
no associated radio emission yet because they are unfavorably
affected by distance, scattering, or lack of sensitive
observations at the time their pulsed radio emission was possibly expected
to be brighter.

The correlation of the potential gap with the X-ray conversion efficiency in magnetars and high-B pulsars (see Figure 2 left) can be interpreted as a natural consequence of the pulsar evolution. To explain this effect, we plot in Figure 2 (right) three evolutionary tracks corresponding to different
magneto-thermal evolutionary models, obtained with the code of \cite{po09}, to which we refer for further details on the physical processes involved.
The values of the dipolar field $B_{\rm p}$ at birth are 
$2\times 10^{13}$, $2\times 10^{14}$, and
$10^{15}$\,G  (internal toroidal components are 0,  $2\times10^{14}$, and $10^{16}$\,G, respectively). These are chosen as representative cases of a high-B pulsar,
a moderate magnetar, and a extreme magnetar (see e.g. Pons \& Perna 2011 for a detailed discussion
of how the luminosity and timing properties depend on the initial field configuration).
We took for all the models a short initial period ($P=0.01$ s), choosing a longer initial period would only shift the lines in the plots towards earlier ages.
From these results, we can conclude that the magneto-thermal-rotational evolution of neutron
stars born with a high magnetic field, say $>5\times 10^{13}$\,G, results in their clustering in a diagonal,
relatively narrow band. Along the evolution, they  cross the $L_{\rm x}/L_{\rm rot}=1$ line very early, 
and spend the rest of their lives in the {\it radio pulsar inactivity} region. 
Typically, an extreme field object crosses the line in less than 1 kyr. This fast motion on the fundamental plane explains
the lack of magnetars in the upper left part of the diagram. Lower field pulsars, on the contrary, 
reach a turning point before the critical line, thus staying in the left side of the diagram where radio pulsar activity is expected.

\section{Discussion}
\label{discussion}

We have shown that radio emitting magnetars have in quiescence $L_{\rm
x}/L_{\rm rot}<1$, as rotation-powered pulsars.  An X-ray efficiency greater than one was all along considered as a basic property to define a magnetar (see e.g. Mereghetti 2008), but this does not hold as such anymore. In magnetars with a small $L_{\rm x}/L_{\rm rot}$ ratio,  particle acceleration and the subsequent ignition of the cascade process could proceed as for normal pulsars \citep{ml10}, successfully reaching the open-field line region and generating pulsed radio emission. This means that their radio emission might basically follow the same rules as for normal radio pulsars, with rotational energy driving pair creation through a cascade, rather than being related to the magnetic energy budget. 

However, while the magnetic field of normal pulsars is dominated by its
dipolar component, in magnetars an important contribution from
higher order multipoles and a toroidal component is expected in both, the internal and external field. This complex magnetic geometry is believed to be ultimately responsible for their bright X-ray emission, their high surface temperature, their bursting and
glitching behavior as well as their outburst activity \citep{td01,tlk02,belo09}. Furthermore, the toroidal geometry is also what is believed to drive the differences in radio properties between
radio magnetars and radio pulsars. In fact, even
if the physical mechanism driving these emissions might
possibly be similar, the actual radio emission appear at a first glance somewhat different. The rapid variability,
broad pulses, and unusually hard radio spectra of magnetars
are consistent with them having a twisted magnetosphere
which is dominated by strongly variable currents
and large plasma densities interfering/interacting with the
pair cascade (Thompson 2008b).
In radio pulsars the plasma density relates to the square
of the emitted radio frequency (Ruderman \& Sutherland
1975). The flat spectra of radio magnetars is then in
line with magnetospheric densities orders of magnitude
larger than in normal pulsars, as theoretically predicted
(Thompson et al. 2002; Nobili et al. 2008) and estimated
from the X-ray spectra (Rea et al. 2008; Zane et al. 2009).
Changes in the magnetospheric density (i.e. during the
outburst decay) might also affect the torque on the neutron
star, as it has been observed in radio-emitting magnetars
(Camilo et al. 2007a), as well as in some radio pulsars
(Kramer et al. 2006).

From Figure 1 it is clear that the high spin-down power is not the only ingredient for magnetars to show or not radio emission: the X-ray luminosity also plays a crucial role (i.e. \sgra\, has a higher $L_{\rm rot}$ than \xte\, or \aa, but it is not radio pulsed; see Figure 1). One possibility to explain the crucial role of the X-ray luminosity is noting that this is mainly related to the toroidal component of the magnetar. In particular, it might be that  radio-emitting magnetars and high-B radio pulsars have systematically lower toroidal fields than the canonical radio-quiet magnetars. This might also be in agreement with the former being fainter, and with a softer X-ray spectrum, during quiescence. A lower crustal toroidal field, in fact, results in less heating produced by Joule dissipation in the star crust, and hence lower surface temperatures. In this picture a possible explanation for the absence of radio emission in the brightest magnetars, is a disruptive interaction between the particle cascades triggered by the acceleration of particles in the electric gap, and the powerful currents forming as a consequence of the largely twisted external magnetic field  \citep{thompson08a,thompson08b,belo09}.

Although the exact relation between the crustal and magnetospheric
$B_{\rm tor}$ components is not known yet, it may be surmised that
stars with a larger internal reservoir of helicity (hence brighter X-ray luminosities) are able to
continuously feed it to the magnetosphere, sustaining a
long-lasting twist. The absence of radio-pulsed emission from
magnetars with high toroidal fields can then be explained
if the particle cascades cannot reach the open-field lines due to
the powerful currents forming as a consequence of the twisted
magnetosphere.

Another possibility might be a reduction in the surface voltage gap
due to pair creation by non-resonant scattering of
high-energy X-rays photons and collisions between gamma-ray and
thermal X-ray photons \citep{thompson08b}. A typical (temperature
dependent) reduction of a factor $\approx 10$--50 in the gap voltage is expected for a
surface temperature $kT>0.1$\,keV. Interestingly, SGRs/AXPs have surface
temperatures $\approx 0.2-0.6$\,keV.

In both these scenarios, the connection of magnetar radio emission with their
X-ray outburst activity is straightforward. 
X-ray outbursts are interpreted as sudden
changes in the magnetic topology, which results in an increase of
the magnetospheric twist, stressing and heating the crust, and
replenishing the magnetosphere with charge \citep{tlk02}.
Around the outburst peak, the twist, the surface temperature and
the magnetospheric charge density attain their largest values. In
this environment, the pair cascades fail to propagate outside
screened by the large currents, and/or the increase in the surface
temperature can reduce drastically the gap voltage. No radio emission is then expected to be detected soon after the outburst onset.

During the outburst decay the magnetosphere untwists, the surface
cools, and radio-pulsed emission may appear. In
particular, the same mechanism which is inhibiting 
radio emission from the canonical magnetars at all times, is also 
responsible for the delay in the activation of radio-emission after the 
onset of an X-ray outburst. Once the radio emission is active again, the
large particle density in the magnetosphere, inheritance of the
increased magnetic twist which caused the outburst, provides an
additional contribution to pair cascade process, hence producing a
much brighter radio emission. As the outburst decays toward
quiescence the magnetosphere progressively untwists,  
the charge density decreases, and the radio flux decays. 

Given the above scenarios, we argue that radio emission may be present at all times in magnetars with $L_{\rm 
x}/L_{\rm rot}<1$,  but during quiescence might be detectable only for 
close objects, while it gets  too dim in other sources (as e.g. \xte).

\section{Conclusions}
\label{conclusions}

In this Letter we discuss the radio emission from magnetars, and its connection with rotation-powered radio pulsars.
Our results can be used to establish if a newly discovered magnetar
will show pulsed radio emission or not. To this end, only the
period, period derivative, and an estimate (or an
upper limit) of its quiescent luminosity are needed (e.g. from available
X-ray all sky surveys). These values will allow to evaluate the electric gap voltage (see Eq. (2)), and the ratio $L_{\rm x}/L_{\rm rot}$. If the electric gap voltage is large enough, and $L_{\rm 
x}/L_{\rm rot}<1$, then radio pulsed emission should be present, and the source eventually detected, if its environment and distance allow such a detection. If the new
source will have  $L_{\rm x}/L_{\rm rot}>1$, it will be radio-quiet, regardless of it showing or 
not X-ray outburst activity.

\acknowledgments
NR is supported by a Ram\'on~y~Cajal fellowship. NR and
DFT  acknowledge support from grants AYA2009-07391, SGR2009-811, TW2010005, and iLINK 2011-0303. JAP acknowledges support from grants AYA2010-21097-C03-02 and
GVPROMETEO2009-103. RT is partially supported by INAF through the funding scheme PRIN 2011. We thank M. Lyutikov and R. Perna for interesting discussions, and A. Possenti, G. L. Israel, P. Esposito and the anonymous referee for their useful comments on the manuscript.

\label{lastpage}


\begin{thebibliography}{}
\bibitem[\protect\citeauthoryear{Anderson et al.}{2011}]{anderson11} Anderson, G., et al. , 2011, ApJ, submitted
\bibitem[\protect\citeauthoryear{Archibald et al. }{2008}]{archibald08} Archibald, A. M., Kaspi, V. M., Livingstone, M. A., McLaughlin, M. A. 2008,  ApJ  688, 550
\bibitem[\protect\citeauthoryear{Baring \& Harding}{1998}]{bh98} Baring, M. G.  \& Harding, A. K., 1998, ApJ, 507, L55
\bibitem[\protect\citeauthoryear{Becker \& Helfand}{1984}]{bh84} Becker, R. H., \& Helfand, D. J. 1984, ApJ, 283, 154
\bibitem[\protect\citeauthoryear{Becker}{2009}]{becker09} Becker, W. 2009, ASSL, 357, Neutron Stars and Pulsars, Springer Berlin Heidelberg, 91 
\bibitem[\protect\citeauthoryear{Beloborodov}{2009}]{belo09} Beloborodov, A. M.  2009, ApJ 703, 1044
\bibitem[\protect\citeauthoryear{Burgay et al.}{2006}]{burgay06} Burgay, M., et al., 2006, MNRAS, 372, 410
\bibitem[\protect\citeauthoryear{Burgay et al.}{2009}]{burgay09} Burgay, M., et al., 2009, ATel, 1913
 \bibitem[\protect\citeauthoryear{Camilo et al.}{2000}]{camilo00}Camilo, F., et al. 2000, ApJ, 541, 367
\bibitem[\protect\citeauthoryear{Camilo et al.}{2006}]{camilo06} Camilo, F., Ransom, S. M., Halpern, J.P., Reynolds, J., Helfand, D. J., Zimmerman, N., Sarkissian, J., 2006, Nature, 442, 892
\bibitem[\protect\citeauthoryear{Camilo et al.}{2007a}]{camilo07a} Camilo, F. et al. 2007a, ApJ, 663, 497
\bibitem[\protect\citeauthoryear{Camilo et al.}{2007b}]{camilo07b} Camilo, F., Ransom, S. M., Halpern, J. P., Reynolds, J. 2007b, ApJ, 666, L93
\bibitem[\protect\citeauthoryear{Camilo et al.}{2008}]{cs08} Camilo, F. \& Sakissian, J. 2008, ATel, 1558
\bibitem[\protect\citeauthoryear{Camilo et al.}{2009}]{camilo09} Camilo, F., Halpern, J. P., Ransom, S. M. 2009, ATel, 1907
\bibitem[\protect\citeauthoryear{Cordes \& Lazio}{2002}]{ne2001} Cordes, J. M. \& Lazio, T. J. W., 2001, astro-ph/0207156
\bibitem[\protect\citeauthoryear{Crawford et al.}{2007}]{crawford07} Crawford, F., Hessels, J. W. T., Kaspi,  V. M. 2007, ApJ, 662, 1183
\bibitem[\protect\citeauthoryear{Esposito et al.}{2009}]{esposito09} Esposito, P.  et al. 2009, MNRAS 399, L44
\bibitem[\protect\citeauthoryear{Gavriil et al.}{2002}]{gavriil02} Gavriil, Kaspi, V. M., Woods, P. M., 2004, Nature, 419, 142 
\bibitem[\protect\citeauthoryear{Gavriil et al.}{2008}]{gavriil08} Gavriil, F. P., Gonzalez, M. E., Gotthelf, E. V., Kaspi, V. M., Livingstone, M. A., Woods, P. M. 2008, Science 319, 1802
\bibitem[\protect\citeauthoryear{Goldreich \& Julian}{1969}]{gj69} Goldreich, P. \& Julian, W. H., 1969,  ApJ,157, 869
\bibitem[\protect\citeauthoryear{Ibrahim et al.}{2004}]{ibrahim04} Ibrahim, A. I.  et al. 2004, ApJ, 609, L21
\bibitem[\protect\citeauthoryear{Kramer et al.}{2006}]{kramer06} Kramer, M., Lyne, A. G., O'Brien, J. T., Jordan, C. A., Lorimer, D. R. 2006, Science 312, 549  
\bibitem[\protect\citeauthoryear{Kouveliotou et al.}{1998}]{kouv98} Kouveliotou, C., et al. 1998 Nature, 393, 235
\bibitem[\protect\citeauthoryear{Kumar \& Safi-Harb}{2008}]{samar08} Kumar, H. S. \& Safi-Harb, S. 2008, ApJ 678, L43
\bibitem[\protect\citeauthoryear{Lazaridis et al.}{2008}]{lazaridis08} Lazaridis, K., et al. 2008, MNRAS 390, 839
\bibitem[\protect\citeauthoryear{Lazarus et al.}{2011}]{lazarus11} Lazarus, P., Kaspi, V. M., Champion, D. J., Hessels, J. W. T., Dib, R. 2011,  ApJ  744, 97
\bibitem[\protect\citeauthoryear{Leahy \& Tian}{2008}]{lt08}  Leahy, D. A., \& Tian, W. W. 2008, A\&A, 480, L25
\bibitem[\protect\citeauthoryear{Levin et al.}{2010}]{levin10} Levin, L.  et al., 2010, ApJ, 721, L33
\bibitem[\protect\citeauthoryear{McLaughlin et al.}{2003}]{mclaughlin03} McLaughlin, M. A., et al. 2003, ApJ, 591, L135
\bibitem[\protect\citeauthoryear{Medin \& Lai}{2010}]{ml10} Medin, Z.  \& D. Lai, D., 2010, MNRAS 406, 1379
\bibitem[\protect\citeauthoryear{Mereghetti}{2008}]{mereghetti08} Mereghetti, S., 2008, A\&AR 15, 225
\bibitem[\protect\citeauthoryear{Ng \& Kaspi}{2011}]{nk11} Ng, C.-Y \& Kaspi, V. M. 2011, Astrophysics of neutron stars 2010: A Conference in Honor of M. Ali Alpar. AIP Conference Proceedings, 1379, 60 
\bibitem[\protect\citeauthoryear{Nobili et al.}{2008}]{ntz08a} Nobili, L., Turolla, R., Zane, S. 2008,  MNRAS 386, 1527 
\bibitem[\protect\citeauthoryear{Pons et al.}{2009}]{po09}  Pons, J. A., Miralles, J. A. \& Geppert, U.  2009, A\&A 496, 207
\bibitem[\protect\citeauthoryear{Pons \& Perna}{2011}]{pp11}  Pons, J. A.  \& Perna, R. 2011, ApJ 741, 123
\bibitem[\protect\citeauthoryear{Rea et al.}{2008}]{rea08} Rea, N., Zane, S., Turolla, R., Lyutikov, M., G\"otz, D., 2008, ApJ 686, 1245
\bibitem[\protect\citeauthoryear{Rea et al.}{2010}]{rea10} Rea, N. et al. 2010, Science, 330, 944
\bibitem[\protect\citeauthoryear{Rea \& Esposito}{2011}]{re11} Rea, N. \& Esposito, P. 2011, APSS, "High-Energy Emission from Pulsars and their Systems", Eds. N. Rea \& D. F. Torres, Springer-Verlag Berlin Heidelberg, 247
\bibitem[\protect\citeauthoryear{Ruderman \& Sutherland}{1975}]{rs75} Ruderman, M. A. \& Sutherland, P. G.,1975,  ApJ, 196 , 51
\bibitem[\protect\citeauthoryear{Serylak et al.}{2009}]{serylak09} Serylak, M., et al. 2009, MNRAS, 394, 295
\bibitem[\protect\citeauthoryear{Thompson \& Duncan}{1995}]{td95} Thompson, C. \& Duncan, R. C., 1995, MNRAS, 275, 255
\bibitem[\protect\citeauthoryear{Thompson \& Duncan}{2001}]{td01} Thompson, C. \& Duncan, R. C., 2001,   ApJ, 561, 980
\bibitem[\protect\citeauthoryear{Thompson et al.}{2002}]{tlk02} Thompson, C., Lyutikov, M., Kulkarni, S. R. 2002, ApJ, 574, 332
\bibitem[\protect\citeauthoryear{Thompson}{2008a}]{thompson08a} Thompson, C., 2008a, ApJ 688, 1258
\bibitem[\protect\citeauthoryear{Thompson}{2008b}]{thompson08b} Thompson, C. 2008b, ApJ 688, 499
\bibitem[\protect\citeauthoryear{Possenti et al.}{2002}]{possenti02} Possenti, A., Cerutti, R., Colpi, M., Mereghetti, S. 2002, A\&A 387, 993
\bibitem[\protect\citeauthoryear{Vink et al.}{2011}]{vby11} Vink, J., Bamba, A.  \& Yamazaki, R.   2011, ApJ 727, 131
\bibitem[\protect\citeauthoryear{Zane et al.}{2009}]{zane09} Zane, S., Rea, N., Turolla, R., Nobili, L. 2009, MNRAS 398, 1403

\end{thebibliography}
\end{document}